\documentclass[%
preprint,
superscriptaddress,
 amsmath,amssymb,
floatfix,
]{revtex4-1}
\usepackage{graphicx}  
\usepackage{ulem}      
\usepackage{lineno}
\usepackage[colorlinks]{hyperref}   

\usepackage{lastpage,xr,refcount,etoolbox}
\begin{document}
\title{Anomalous temperature dependence of optical and acoustic phonons in Bi$_2$Se$_3$ arising from stacking faults}

\author{Gyan Prakash} \affiliation{Department of Physics and Center for Ultrafast Laser Applications, Indian Institute of Science, Bangalore 560012, India}

\author{Koushik Pal} \affiliation{Theoretical Sciences Unit, Jawaharlal Nehru Centre for Advanced Scientific Research,   Jakkur campus, Bangalore 560064}

\author{U. V. Waghmare} \affiliation{Theoretical Sciences Unit, Jawaharlal Nehru Centre for Advanced Scientific Research, Jakkur campus, Bangalore 560064}

\author{A. K. Sood}
\affiliation{Department of Physics and Center for Ultrafast Laser Applications, Indian Institute of Science, Bangalore 560012, India}

\begin{abstract}  
The class of layered 3D-topological insulators  have shown intriguingly anomalous temperature dependence in their thermal expansion coefficients. It was  proposed that stacking faults are the origin of the observed anomalous thermal expansion.   Here, using  femtosecond pump-probe differential reflectivity measurements we probe the carrier and coherently generated totally symmetric A$_{1g}$  optical phonons in Bi$_2$Se$_3$. Transient signals  also show a low frequency ($\sim$GHz) oscillations due to coherent longitudinal acoustic phonons. We extract temperature dependence of optical constants, sound velocity  and Young's modulus    of  Bi$_2$Se$_3$   using the strain pulse propagation model. A remarkable anomalous behavior around $\sim$180 K is observed in the temperature dependence of optical and acoustic phonons as well as the  optical constants.  First-principles density functional theory (DFT) reveals that thermally activated formation of stacking faults is responsible for the anomalies observed in Bi$_2$Se$_3$, similar to case of Sb$_2$Te$_3$. We also show that inclusion of spin-orbit coupling plays an important role in reducing the total energy difference between the pristine and the faulted structures.

\end{abstract}
\pacs{}
\maketitle
\section{\label{sec:level1}  Introduction}
 Many compounds (e.g., Bi$_2$Se$_3$ , Bi$_2$Te$_3$,
hexagonal Sb$_2$Se$_3$, $\beta$-As$_2$Te$_3$)  in the group
V-VI chalcogenide family have been shown to exhibit strong topological insulating phase \cite{xia2009observation, bera2013sharp, pal2014strain} as well as excellent thermolectric properties \citep{poudel2008high, caillat1992thermoelectric, wright1958thermoelectric, hinsche2012thermoelectric, Thonhauser}. Hence, understanding the temperature dependence of structural,
vibrational and  electronic properties of  these compounds are of paramount importance. Among the above
compounds, Sb$_2$Te$_3$ \cite{Dutta, chen2011thermal}, Bi$_2$Se$_3$ \cite{chen2011thermal} and Bi$_2$Te$_3$ \cite{Li} were shown to exhibit an 
intriguing temperature dependent anomaly  in their linear thermal expansion coefficient along the trigonal $c$ axis ($\alpha_\|$). The coefficient  $\alpha_\|$ exhibits a minimum at a certain temperature $T_a$ ($T_a \sim$225 K for Sb$_2$Te$_3$, $\sim$ 180 K for Bi$_2$Se$_3$ and $\sim$240 K for Bi$_2$Te$_3$). Recently, our group \citep{Prakash} extensively studied the electron and phonon dynamics of Sb$_2$Te$_3$ 
by means of ultrafast pump-probe spectroscopy to shed light on the mechanisms responsible behind the
observed  structural and vibrational anomalies.   In our earlier temperature dependent transient differential reflectivity study on Sb$_2$Te$_3$ we observed anomalous behavior in the relaxation dynamics of the photoexcited carriers, optical phonons and acoustic phonons corresponding to temperature associated with the anomalous thermal expansion \citep{Prakash}. Therein, we showed  stacking fault as the  origin of the anomaly and proposed that the same mechanism may be responsible for other materials in this class. The aim of the present work is to see if this is indeed the case or not. We take the case of Bi$_2$Se$_3$ from the same family of layered 3D TI and performed a combination of experimental and theoretical investigations. Using ultrafast pulses of photon energy $\sim$ 1.57 eV, in a pump-probe experiment,  we probe the carrier and phonon dynamics in Bi$_2$Se$_3$  in the temperature range 45 to 300 K with particular emphasis on exploring the effect of anomaly in the thermal expansion coefficient.  

Pump induced electronic and thermal strain at the surface generates acoustic phonons \citep{thomsen1986surface, Prakash, Kumar}.  Differential reflectivity signals of Bi$_2$Se$_3$ show clear oscillations related to acoustic phonons. Temperature dependent acoustic phonon frequency and decay time contain valuable information about structural transitions.  Using strain pulse propagation model we extract the temperature dependence of the optical constants ($n$ and $k$), sound velocity ($v_{_{LA}}$) and Young's modulus of elasticity ($Y_{_{LA}}$).  The obtained results show clear anomalous behavior of optical and acoustic phonons as well as  the  optical constants ($n$ and $k$)  around $T_a \sim \text{180 K}$.   Using first-principles density functional theory (DFT) calculations, we investigate the origin of the observed anomalies through introducing stacking faults in the crystal structure of Bi$_2$Se$_3$. Our results confirm the formation of the stacking fault as the generic mechanism responsible for the observed anomaly in various parameters in this class of materials. Spin-orbit coupling (SOC) is known to play an important role in strong topological insulators. Our results show that the SOC not only affects electronic topology, but also the energetics of different structures of these TIs. Inclusion of SOC reduces the total energy difference (at 0 K) between the pristine and the faulted structures. This total energy difference at 0 K (obtained with SOC) when combined with the difference of their vibrational contribution to the free energy, predicts a  temperature of stabilization of stacking faults closer to the experimentally observed $T_a$.

\section{\label{sec:level2}  Experimental Details}
Temperature dependent ultrafast differential reflectivity measurements were done in a continuous helium flow cryostat using pump and probe photons of energy 1.57 eV and pulse duration $\sim$60 fs from a regenerative amplifier (Spitfire from Spectra-Physics) operating at 1 kHz repetition rate. Single crystal of Bi$_2$Se$_3$  was transferred  into the evacuated cryostat immediately after cleaving the top surface to avoid any surface contamination. Pump-probe measurements were done with orthogonal polarized pump and probe beams with spot sizes of 850 $\mu$m and 130 $\mu$m at the overlap position, respectively. Analyzers orthogonal to pump beam were used to avoid pump scattering.  The signal to noise ratio was improved by phase locked detection where the pump beam was modulated with a mechanical chopper and a lock-in-amplifier was slaved to this frequency.  Pump and probe fluences were kept at $\sim$ 620 $\mu$J/cm$^2$ and 70 $\mu$J/cm$^2$ (much below the damage threshold of Bi$_2$Se$_3$), respectively  for all the temperature dependent differential reflectivity measurement\citep{hsieh2011nonlinear}.
     
\section{\label{sec:level3}  Results and Discussion}
\subsection{Experimental Analysis} 
\begin{figure*}
\begin{center}
\includegraphics[width=0.7\linewidth]{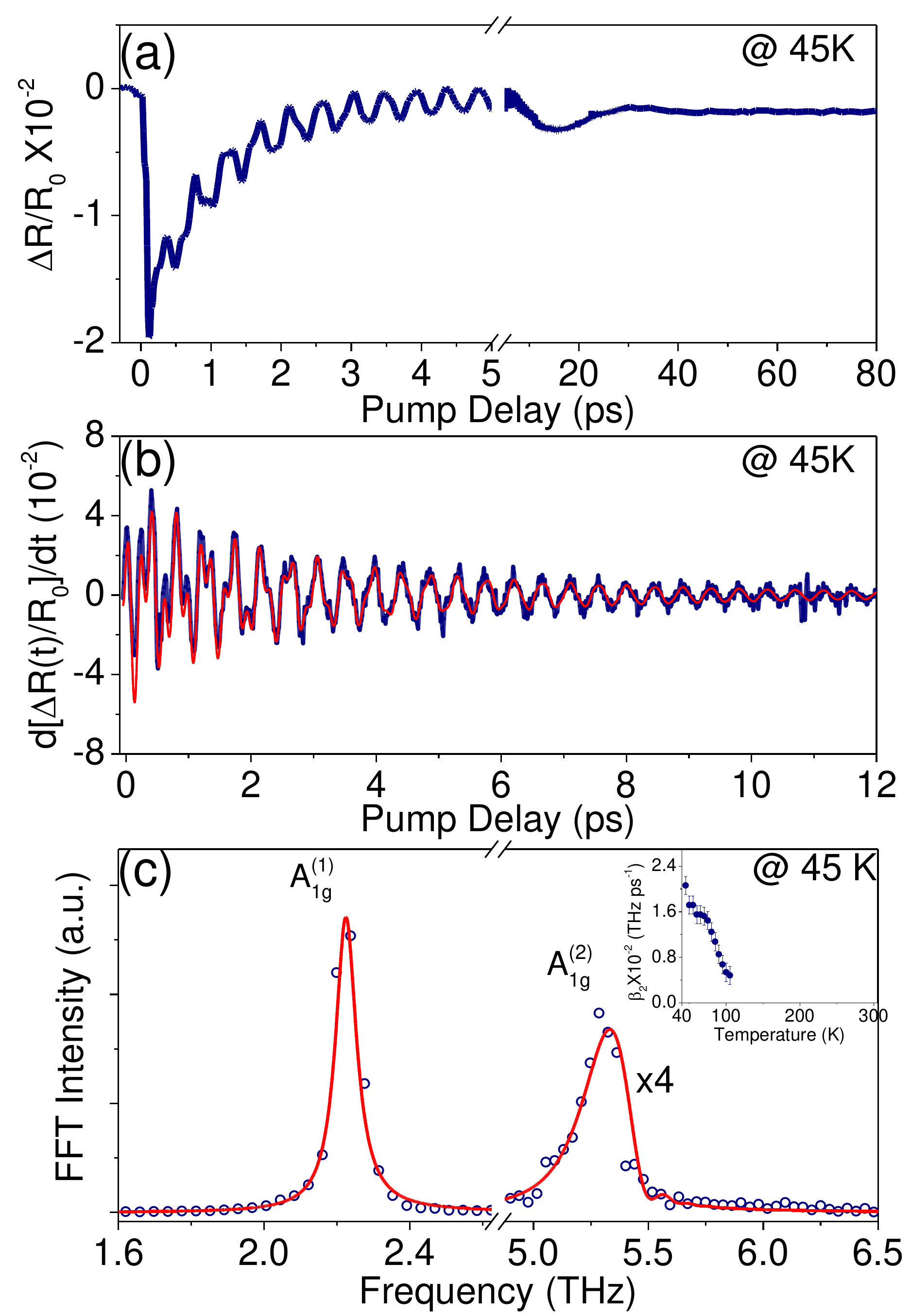}

\caption{\textbf{Bi$_2$Se$_3$ pump-probe signal at 45 K} (a) $\Delta \text{R}/\text{R}_0$ signal at 45 K showing transient oscillations due to two coherent optical phonons (A$_{1g}^{(1)}$ and A$_{1g}^{(2)}$) and an acoustic phonon over a background of electronic relaxation. (b) Time derivative of the pump-probe signal in (a). Solid line in red is fit to the experimental data. (c) FFT of the time derivative data and the fit in (b). Inset shows temperature dependence of the chirp parameter $\beta_2$.} 
\label{fig:TD_DataFFTBi2Se3}
\end{center}
\end{figure*}

The layered crystal structure of Bi$_2$Se$_3$ is  made up of close-packed atomic layers which are periodically stacked along the c-axis with a unit of five atomic planes (Se$^{(1)}$-Bi-Se$^{(2)}$-Bi-Se$^{(1')}$)  called quintuple belonging to the space group symmetry $\text{R}\bar{3}m$ (No 166).  Group theoretical analysis of the symmetry of centrosymmetric rhombohedral ($D_{3d}^5$)  crystal predicts 12 optical phonons at the $\Gamma$ point with irreducible representations as $2A_{1g}$(R)$+2E_g$(R)$+2A_{2u}$(IR)$+2E_u$(IR) \citep{richter1977raman}, where R and IR refer to Raman and infrared active modes, respectively \citep{richter1977raman}. Fig. \ref{fig:TD_DataFFTBi2Se3} (a) shows the differential reflectivity signal $\Delta \text{R}(t)/\text{R}_0$  obtained following pump excitation at $\text{45 K}$. Transient kinetics clearly shows two superimposed oscillations of THz frequency corresponding to the coherent optical phonons. The frequencies of these superimposed  oscillations are due to the displacive excitation \citep{zeiger1992theory, kamaraju2010temperature} of A$_{1g}^{(1)}$ and A$_{1g}^{(2)}$  optical phonon modes, confirmed by their oscillation frequencies of 2.23 THz and 5.29 THz  observed at $\text{45 K}$, respectively \citep{richter1977raman, kim2012temperature}. As can be clearly seen from the transient $\Delta \text{R}(t)/\text{R}_0$ signal in Fig. \ref{fig:TD_DataFFTBi2Se3}(a), the low frequency mode A$_{1g}^{(1)}$ is long lived and is observed upto $\sim$ 12 ps.  The high frequency mode A$_{1g}^{(2)}$ is however, short lived and is observed upto $\sim$ 5 ps only. A slow oscillation of GHz frequency together with an exponentially decaying background of carrier relaxation is also observed in the differential reflectivity signal of Bi$_2$Se$_3$. The slow oscillation observed is due to the coherent acoustic phonon \citep{wu2007femtosecond, thomsen1986surface}.

     As the higher frequency mode A$_{1g}^{(2)}$ is short lived, the amplitude, scattering rate,  frequency and chirp  of this mode are obscured by the underlying  signal due to the electronic relaxation and the coherent acoustic phonon.   To clearly observe and extract the temperature dependence of the physical parameters associated with the coherent optical phonons A$_{1g}^{(1)}$ and A$_{1g}^{(2)}$ in Bi$_2$Se$_3$, we remove  the signal originating from the electronic relaxation and the slow oscillation due to the coherent acoustic phonons by numerically differentiating the $\Delta \text{R}(t)/\text{R}_0$ data with respect to delay time at each temperature \citep{kamaraju2011coherent, kamaraju2010temperature, wang2008reduction, wu2008ultrafast}.  The electronic and coherent acoustic phonon relaxations are analysed separately and discussed later. The obtained  transient signal after taking the time derivative is proportional to the normal phonon coordinate $Q_0(t)$ of the coherent optical phonons \citep{kamaraju2010temperature, wang2008reduction, wu2008ultrafast}. A representative plot showing the time derivative of the $\Delta \text{R}/\text{R}_0$ signal at 45 K and its fast Fourier transform (FFT) are displayed in Fig. \ref{fig:TD_DataFFTBi2Se3} (b) and (c), respectively. The FFT intensity of the A$_{1g}^{(1)}$ mode exhibits a symmetric Lorentzian lineshape. However, in contrast to the A$_{1g}^{(1)}$ mode the lineshape of the A$_{1g}^{(2)}$ mode is asymmetric. The asymmetric lineshape indicates the presence of chirp in the A$_{1g}^{(2)}$ mode \citep{kamaraju2010temperature, wang2008reduction, wu2008ultrafast}.
   
   We model the time derivative signal as sum of two chirped damping harmonic oscillators \citep{wu2008ultrafast, kamaraju2010temperature}

\begin{equation}
 \frac{d (\Delta \text{R}/\text{R}_0)}{dt} =   \sum_{i=1,2} \text{B}_i \exp({-\Gamma_i t}) \cos\left[2 \pi (\nu_i +\beta_i t) t+\phi_i \right]
 \label{Eq:OsciTimeDervBi2Se3}
 \end{equation}

\begin{table*}[ht]
\caption{\label{tab:tableBi2Se3} The fitting parameters $\nu_{i0}$, $A_i$, $C_i$ and $\Gamma_{i0}$ used in fitting the temperature dependence of scattering rates and frequencies of  Bi$_2$Se$_3$ coherent optical phonons   A$_{1g}^{(1)}$ and A$_{1g}^{(2)}$. }

\begin{tabular}{c c c c c c}
\hline
Optical phonon mode &  $\nu_{i0}$ (THz) & $A_i$ (THZ) & $C_i$ (THZ) &  $\Gamma_{i0}$ (THz) \\

\hline
($i$=1) A$_{1g}^{(1)}$  &2.237$\pm$0.001 & -0.0038$\pm$0.0001 & 0.017$\pm$0.001    &  0.149$\pm$0.001 \\
($i$=2) A$_{1g}^{(2)}$  &5.49$\pm$0.01 & -0.001$\pm$0.001 & 0.17$\pm$0.01    &  0.32$\pm$0.01 \\

\hline
\end{tabular}
\end{table*}

 where $\text{B}_i$, $\Gamma_i$, $\nu_i$ and  $\phi_i$ are the amplitude, scattering rate, frequency and initial phase of the coherent optical phonons A$_{1g}^{(1)}$ and A$_{1g}^{(2)}$, respectively. The parameter $\beta_i$ is the chirp parameter describing the linear sweep in the phonon frequency with the pump-probe delay time. The indices $i$=1 and $i$=2 correspond to the optical phonon A$_{1g}^{(1)}$ and A$_{1g}^{(2)}$, respectively. Eq. (\ref{Eq:OsciTimeDervBi2Se3}) furnishes excellent fit to the time derivative data at all temperatures. The A$_{1g}^{(1)}$ optical phonon did not show chirp ($\beta_1=0$)  at all temperatures.  The chirp in the A$_{1g}^{(2)}$ optical phonon was found to be highest at 45 K ($\beta_2=2\times 10^{-2}$ THz ps$^{-1}$), which  decreased as the sample temperature was raised and completely disappeared for  temperatures above 105 K (inset of Fig. \ref{fig:TD_DataFFTBi2Se3} (c)).  A fit to the time derivative data at 45 K is shown in panel (b) of Fig. \ref{fig:TD_DataFFTBi2Se3}.  As shown in Fig. \ref{fig:TD_DataFFTBi2Se3}(c) the FFT of fitted curve in time-domain  also captures the dynamics of the A$_{1g}^{(1)}$  and A$_{1g}^{(2)}$ modes in frequency-domain.

\begin{figure*}[ht]
\centering
\includegraphics[width=0.8\linewidth]{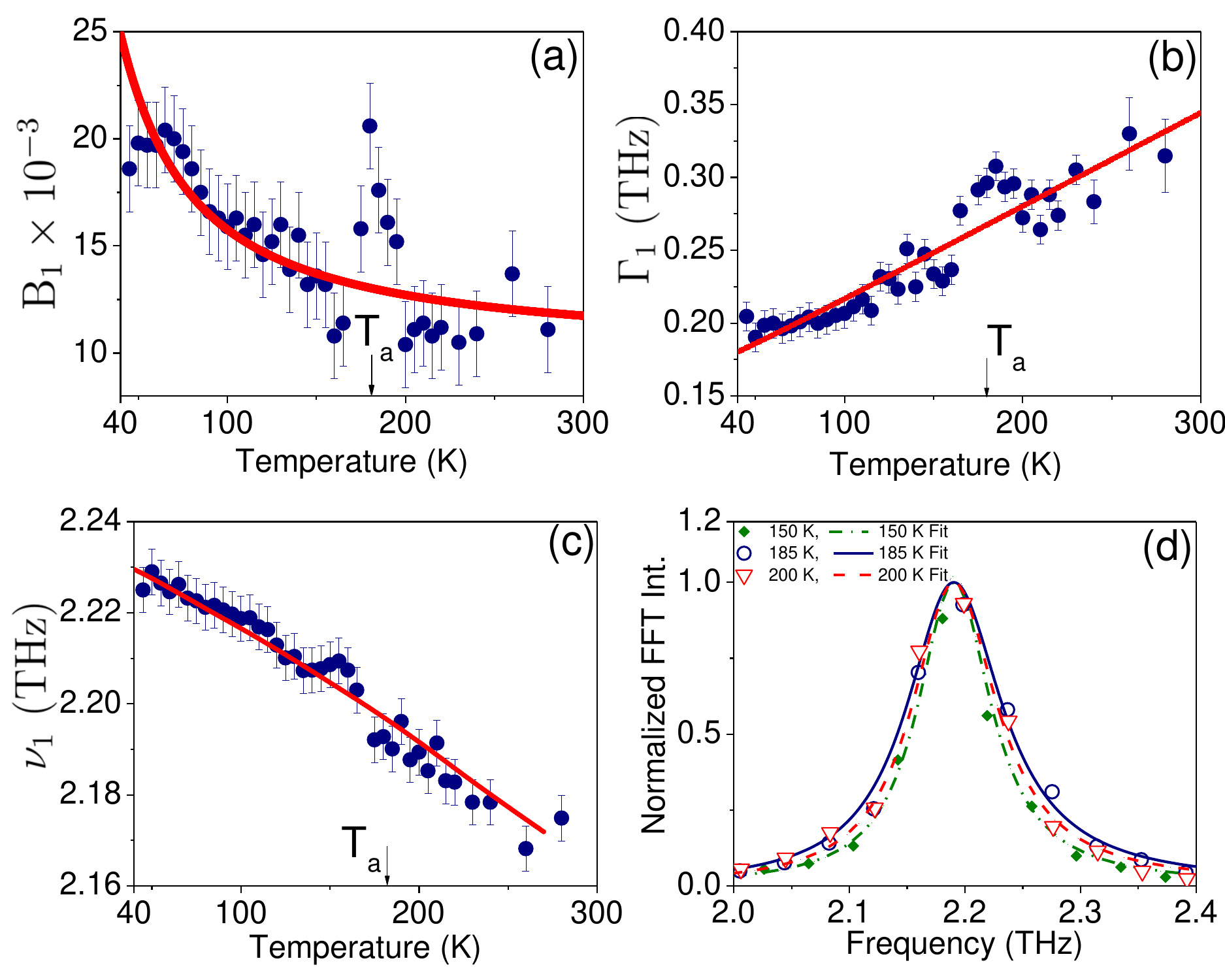}

\caption{\textbf{Bi$_2$Se$_3$ coherent optical phonon A$_{1g}^{(1)}$.} Temperature dependence of the (a) amplitude (b) scattering rate and (c) frequency shift of the Bi$_2$Se$_3$ coherent optical phonon A$_{1g}^{(1)}$. Solid line in (a) and (b) are fit to the data using cubic anharmonic model.  Solid line in (c) is fit to the frequency shift including quasi-harmonic contribution along with the cubic anharmonic contribution. (d) FFT of the time derivative signal at  temperatures 150 K, 185 K and 200 K for comparing the linewidth increase around  the anomaly temperature T$_a \sim$180 K. Frequencies of 150 K and 185 K plots are shifted to match FFT intensity maximum  peak of the plot at  200 K for better comparison of the linewidth.} 
\label{fig:A1g1_Bi2Se3}
\end{figure*}

\begin{figure*}[ht]
\begin{center}
\includegraphics[width=0.8\linewidth]{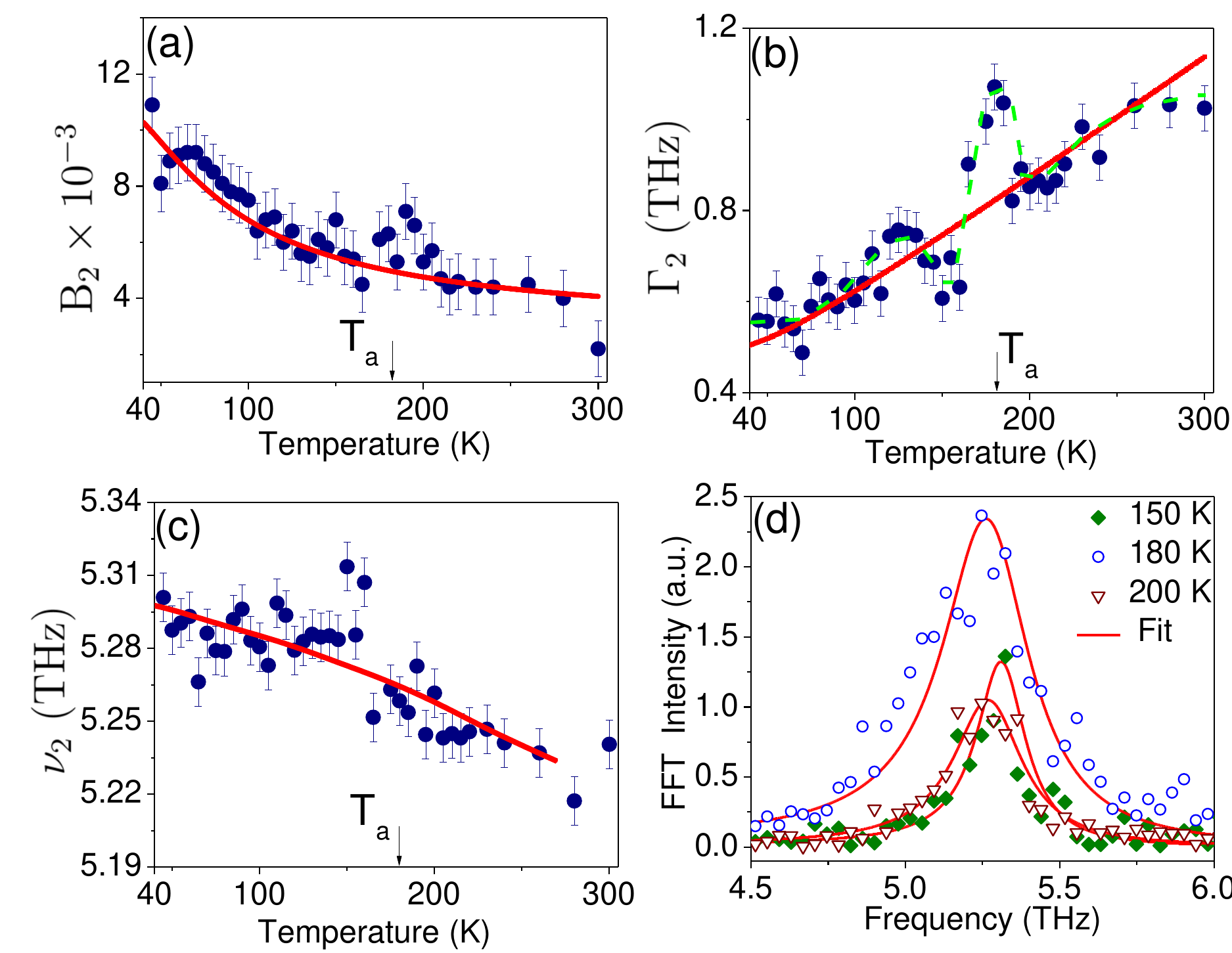}
\caption{\textbf{Bi$_2$Se$_3$ coherent optical phonon A$_{1g}^{(2)}$.} Temperature dependence of the (a) amplitude (b) scattering rate and (c) frequency shift of the Bi$_2$Se$_3$ coherent optical phonon A$_{1g}^{(2)}$. Solid line in (a) and (b) are fit to the data using cubic anharmonic model.  Solid line in (c) is fit to the frequency shift including quasi-harmonic contribution along with the cubic anharmonic contribution. Dashed line in (b) through the data points in green color is guide to the eye. (d) FFT of the time derivative signal at three temperatures 150 K, 180 K and 200 K for comparing the linewidth increase around the anomaly temperature T$_a \sim$180 K. Solid line in red is the FFT of the fit to the time derivative signal in time-domain. } 
\label{fig:A1g2_Bi2Se3}
\end{center}
\end{figure*}

 The optical phonon parameters obtained from the fit at all temperatures are shown in Fig. \ref{fig:A1g1_Bi2Se3} (a-c) and Fig. \ref{fig:A1g2_Bi2Se3} (a-c).   We fit the amplitudes $\text{B}_i$ (Fig. \ref{fig:A1g1_Bi2Se3} (a) and  \ref{fig:A1g2_Bi2Se3} (a)) and scattering rates  $\Gamma_i$ (Fig.  \ref{fig:A1g1_Bi2Se3} (b) and \ref{fig:A1g2_Bi2Se3} (b)) of  A$_{1g}^{(1)}$  and A$_{1g}^{(2)}$ optical phonons with functions   $\text{B}_i  \sim \frac{n\left(\nu_{i0}\right)+1}{2n\left(\nu_{i0}/2 \right)+1}$ and $\Gamma_i (\text{T})= \Gamma_{i0}+C_i\left[2n(\nu_{i0})+ 1\right]$ based on simple cubic anharmonic decay model \citep{klemens1966anharmonic}, respectively. The parameter $\Gamma_{i0}$ is the disorder-induced temperature-independent decay rate, $C_i$ is related to phonon-phonon interaction and $\text{n}(\nu_{i0})=\exp \left[\left(\frac{h \nu_{i0}}{k_BT}-1\right)\right]^{-1}$ is the Bose-Einstein statistical factor with indices $i$=1 and $i$=2 corresponding to A$_{1g}^{(1)}$ and A$_{1g}^{(2)}$ optical phonons with frequencies $\nu_{10}$ and $\nu_{20}$, respectively.   It can be seen that the  amplitudes of the A$_{1g}^{(1)}$ and A$_{1g}^{(2)}$ optical phonons  are anomalously high near the anomaly temperature $\text{T}_a$ $\sim$ 180 K (A$_{1g}^{(1)}$ amplitude is $\sim$58$ \%$ and A$_{1g}^{(2)}$ amplitude is $\sim$45$ \%$ higher than the value expected from the cubic anharmonic fit). The scattering rates of both A$_{1g}^{(1)}$ and  A$_{1g}^{(2)}$ optical phonons are also higher than the scattering rates predicted by the cubic anharmonicity model near the anomaly temperature $\text{T}_a$ $\sim$ 180 K. The increased scattering rate is also confirmed by the accompanied linewidth increase clearly seen in the FFT intensity of the   A$_{1g}^{(1)}$ and A$_{1g}^{(2)}$ optical phonon, as shown in Fig. \ref{fig:A1g1_Bi2Se3} (d)  and \ref{fig:A1g2_Bi2Se3}(d). The enhanced phonon scattering rates ($\Gamma_1$ and $\Gamma_2$) near T$_a$ is similar to the case of Sb$_2$Te$_3$ \citep{Prakash}, arising from softness of the lattice in the temperature range near T$_a$.

The temperature dependent frequency shift in simple cubic anharmonic model is given by $\nu_i(T)=\nu_{i0}+A_i \left[2 \text{n}(\nu_{i0})+1\right]$ where $A_i$ is related to the phonon-phonon interaction. A more accurate description to temperature dependent frequency shift is given by including the quasi-harmonic contribution due to volume expansion \citep{kim2012temperature, zouboulis1991raman,   borer1971line, postmus1968pressure}: $\nu_i(\text{T})=\nu_{i0}+A_i\left[2n(\nu_{i0})+ 1\right]+\nu_{i0}\left[\exp\left(-\gamma\int_0^{\text{T}}\left[\alpha_\|\left(\text{T}'\right)+2\alpha_\perp\left(\text{T}'\right)\right]  d\text{T}'\right)-1\right]$ where, $\gamma$ is mode Gr$\ddot{\text{u}}$iensen parameter,  $\alpha_\|$ and $\alpha_\perp$ are the coefficients of thermal expansion along the `a' and `c' axis of the hexagonal lattice, respectively. We fit the frequency shift of  A$_{1g}^{(1)}$ and  A$_{1g}^{(2)}$ with temperature (Fig. \ref{fig:A1g1_Bi2Se3} (c) and Fig. \ref{fig:A1g2_Bi2Se3} (c)) by including the contribution due to cubic anharmonicity as well as quasi-harmonic. To capture the effect of anomalous thermal expansion on phonon frequencies, we used the experimental values of $\alpha_\parallel$(T) and $\alpha_\perp$(T) from Ref. [30] and mode Gr$\ddot{\text{u}}$einsen parameter $\gamma$=1.4 \citep{chen2011thermal} to calculate the quasi-harmonic contribution to frequency shift of both A$_{1g}^{(1)}$ and A$_{1g}^{(2)}$ optical phonons. The resulting fits are shown in Fig. \ref{fig:A1g1_Bi2Se3}(c) and \ref{fig:A1g2_Bi2Se3}(c). Table (\ref{tab:tableBi2Se3}) summarizes the fit parameters. It can be seen that the temperature dependence of A$_{1g}^{(1)}$  and A$_{1g}^{(2)}$ mode frequencies,  well captured by  the anharmonic model, do not show a clear decrease in phonon frequency near T$_a$, probably due to larger error bars.

\begin{figure*}[ht]
\begin{center}
\includegraphics[width=0.8\linewidth]{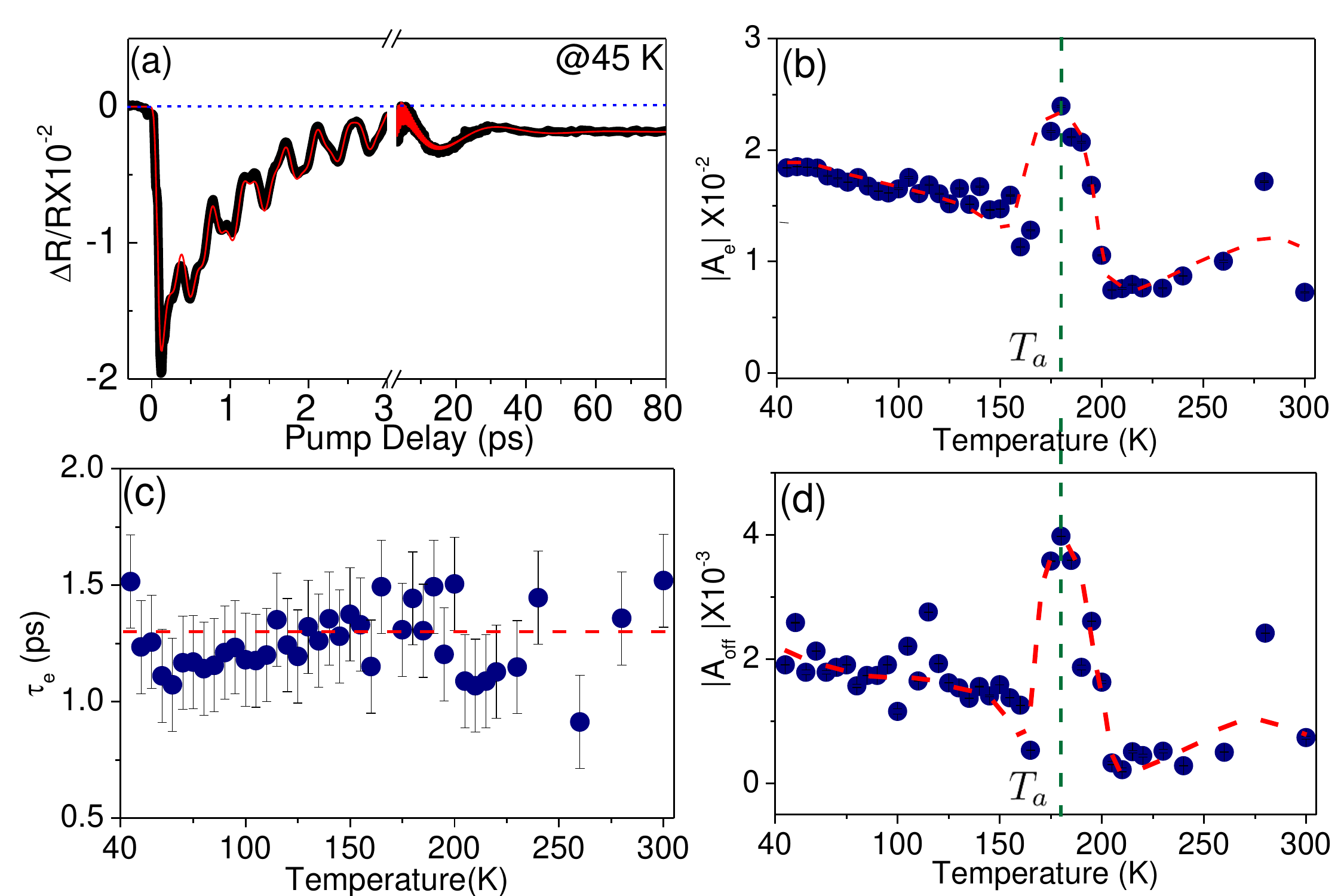}

\caption{\textbf{Electronic relaxation Bi$_2$Se$_3$} (a) $\Delta \text{R}/\text{R}_0$ signal at 45 K with fit to the experimental data as discussed in the text. Temperature dependence of electronic  relaxation parameters obtained from the fit to the experimental $\Delta \text{R}/\text{R}_0$ signal: (b) amplitude, (c) relaxation time of electron-phonon mediated relaxation of photoexcited carriers and (d) amplitude of long time decaying background level.  Dashed lines joining data points in (b), (c) and (d) are guide to eye. Error bars in (b) and (d) are smaller than the symbol size.} 
\label{fig:RepFit_ElecComp_Bi2Se3}
\end{center}
\end{figure*}

To extract the information about the temperature dependence of the photoexcited carriers and the coherent acoustic phonons in Bi$_2$Se$_3$, we fit the $\Delta \text{R}(t)/\text{R}_0$ data with the following equation convoluted with a Gaussian laser pulse, at each temperature:

\begin{eqnarray}
\label{Eq:ElectAndAcousticPhononBi2Se3}
\frac{\Delta \text{R}(t)}{\text{R}_0}&=&  \text{A}_{e} \exp(-t/\tau_{_e}) \nonumber\\&& +\sum_{i=1,2} \text{B}_i \exp({-\Gamma_i t}) \sin\left[2 \pi (\nu_i +\beta_i t) t+\phi_i \right] \text{ }\\&& +\text{B}_{_{LA}} \exp({-t/\tau_{_{LA}}}) \cos(2\pi\nu_{_{LA}} t+\phi_{_{LA}}) +\text{A}_{off}   \nonumber
\end{eqnarray}

where A$_e$ and $\tau_e^{-1}$ are the amplitude and electron-phonon mediated relaxation rate of photoexcited carriers, respectively. A$_{off}$ is the unrecovered background associated with the diffusion of localized photoexcited carriers and their recombination \citep{glinka2014effect}.  $\text{B}_i$ ($\text{B}_{LA}$), $\Gamma_{i}$ ($\tau_{LA}$), $\nu_{i}$ ($\nu_{LA}$) and $\phi_{i}$ ($\phi_{LA}$) are the amplitude, scattering rate (decay time), frequency and the phase of the coherent optical (acoustic)  phonons, with the indices $i=1, 2$ and $LA$  corresponding to  optical phonon A$_{1g}^{(1)}$, A$_{1g}^{(2)}$ and the longitudinal acoustic phonon, respectively. $\beta_i$ is the chirp parameter of the optical phonons.  While fitting the $\Delta \text{R}/\text{R}_0$ signal the optical phonon parameters $\Gamma_{i}$, $\nu_{i}$, $\phi_i$ and $\beta_{i}$ were fixed to the values obtained by fitting the time-derivative signal (Eq. 1). The amplitude $\text{B}_i$ was scaled to get the best fit.  Fig. \ref{fig:RepFit_ElecComp_Bi2Se3} (a) shows the fit to the experimental data at 45 K with Eq. (2). The temperature dependence of the prameters related to the carrier relaxation   A$_e$,  $\tau_e^{-1}$ and A$_{off}$ are shown in Fig. \ref{fig:RepFit_ElecComp_Bi2Se3} (b-d). The electronic relaxation time ($\tau_e$)  is $\sim$1.5 $\pm$0.3 ps at 300 K, similar to the earlier reported results in Bi$_2$Se$_3$ thin films \citep{glinka2014effect}. As shown in Fig. \ref{fig:RepFit_ElecComp_Bi2Se3} (c) $\tau_e$ does not show any variation with temperature.   However, the amplitudes  $\vert \text{A}_e \vert$ and $\vert \text{A}_{off} \vert$ show a significant increase around $\sim$180 K (Fig. \ref{fig:RepFit_ElecComp_Bi2Se3} (b) and (d)), similar to the results in Sb$_2$Te$_3$ \citep{Prakash}. The changes in $\vert \text{A}_e \vert$ and $\vert \text{A}_{off} \vert$ are related to the temperature dependent reflectivity, discussed as follows.

\begin{figure*}
\begin{center}
\includegraphics[width=0.8\linewidth]{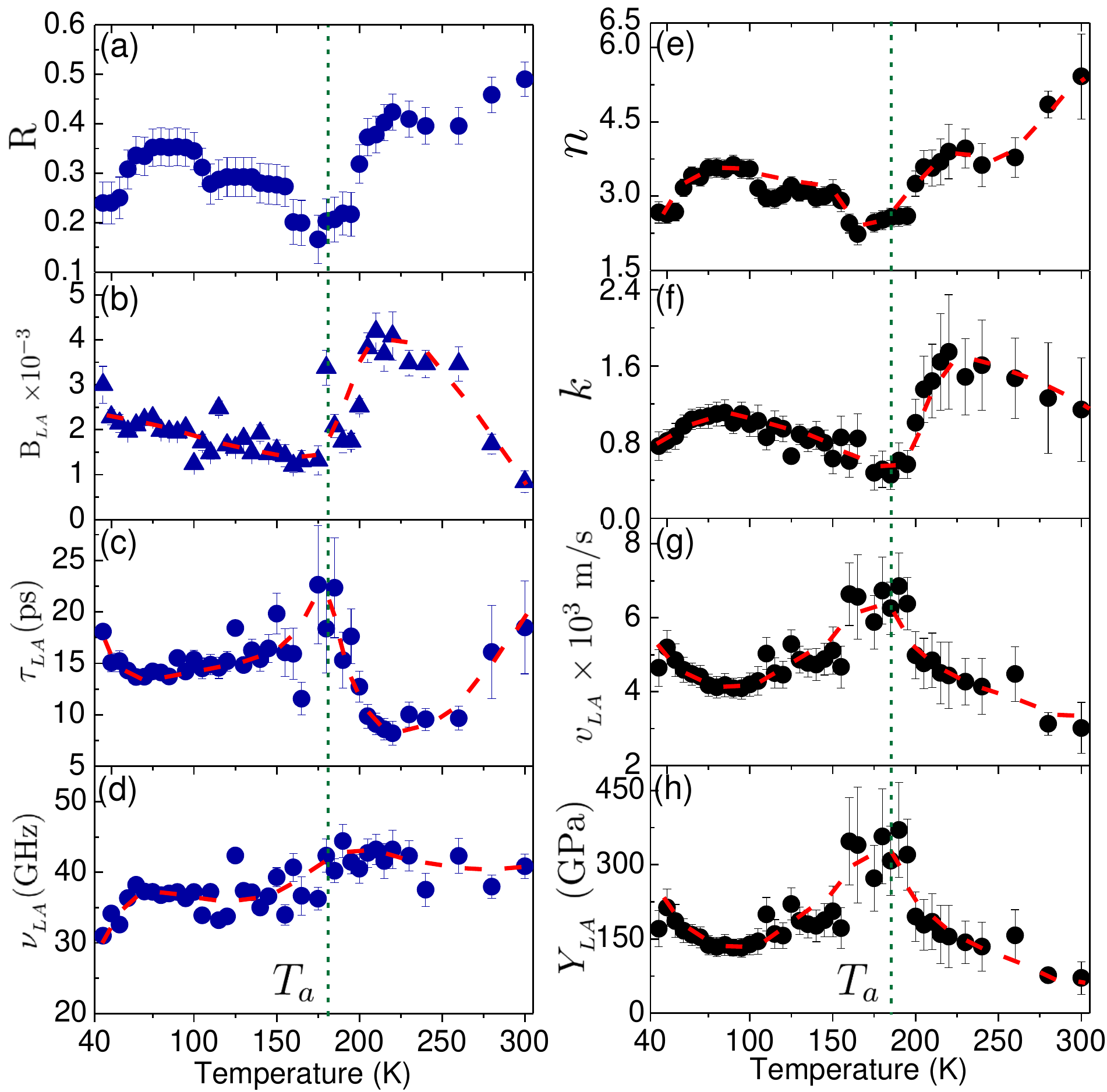}

\caption{\textbf{Bi$_2$Se$_3$ coherent longitudinal acoustic phonon.}  (a) Temperature dependence of  reflectivity of unexcited Bi$_2$Se$_3$ at 1.57 eV. (b) Acoustic phonon amplitude ($\text{B}_{_{LA}}$), (c) dephasing time ($\tau_{_{LA}}$), (d) frequency ($\nu_{_{LA}}$), (e and f) optical constants  ($n$ and $k$), (g) sound velocity $v_{_{LA}}$ and (h) Young's modulus of elasticity ($Y_{_{LA}}$). Dashed lines joining through the data points are guide to eye.}
\label{fig:AcousPhononBi2Se3}
\end{center}
\end{figure*}

Fig. \ref{fig:AcousPhononBi2Se3} (b), (c) and (d) shows the amplitude, decay time and the frequency of the coherent acoustic phonon obtained by fitting $\Delta \text{R}/ \text{R}_0$  with Eq. (\ref{Eq:ElectAndAcousticPhononBi2Se3}). Reflectivity of the Bi$_2$Se$_3$ sample  at 1.57 eV, shown in Fig. \ref{fig:AcousPhononBi2Se3} (a), was obtained by recording the probe intensity keeping the pump beam blocked. We calculated the optical constants $n$ and $k$, sound velocity ($v_{LA}$) and Young's modulus of elasticity ($Y_{LA}$) from the experimental acoustic phonon parameters and the reflectivity using strain pulse propagation model  \cite{Prakash, thomsen1986surface, Kumar}. The procedure adopted is as follows.  The change in probe reflectivity  owing to strain pulse is given by \citep{thomsen1986surface, Prakash}

\begin{eqnarray}
\Delta \text{R}\propto \cos{\left({\frac{4\pi n v{_{LA}}t}{\lambda}}-\delta\right)}e^{-z/\xi}
\end{eqnarray}

where $\delta $ is the phase shift, $v_{LA}$ sound velocity, $n$ refractive index, $\xi=\lambda/4\pi k$  is probe absorption length and $\lambda$ is the probe wavelength. The longitudinal acoustic phonon frequency is given by  $\nu_{LA}=2nv_{LA}/\lambda$. The damping time is related to the duration the strain pulse takes to travel a distance equal to the penetration length of the probe pulse, $\tau_{LA}=\xi/v_{LA}$. From above relations we have $n/k=2 \pi \nu_{LA} \tau_{LA}$. For  normal incident light, reflectivity is given by R(T)=$\left[(n-1)^2+k^2\right]/\left[(n+1)^2+k^2\right]$. Using $\text{R}(\text{T})$, $\nu_{LA}$(T) and $\tau_{LA}$(T) we  obtain refractive index $n(\text{T})$ and $k(\text{T})$. The longitudinal modulus of elasticity related to LA modes along the c-axis is given by 
$\text{Y}_{LA}(\text{T})=v_{LA}^2({\text{T}}) \rho(\text{T})$, where $\rho(\text{T})$ is density calculated using temperature dependence of the 
lattice parameters of Bi$_2$Se$_3$ \cite{chen2011thermal}. The temperature dependence of physical parameters  $n$ and $k$, $ v_{LA}$ and $Y_{LA}$  obtained  from  strain pulse propagation analysis  are shown in Fig. \ref{fig:AcousPhononBi2Se3} (e-h).  We see that near the anomaly tmeperature $\text{T}_a \sim$180 K there is a drop in the optical constants ($n$ and $k$) (Fig. \ref{fig:AcousPhononBi2Se3} (e) and (f)) and  a anomalous increase in the sound velocity ($ v_{_{LA}}$) and Young's modulus ($Y_{_{LA}}$) (Fig. \ref{fig:AcousPhononBi2Se3} (g) and (h)) similar to that observed in Sb$_2$Te$_3$ \citep{Prakash}.

\subsection{Theoretical Analysis} 

To understand the physical mechanism responsible behind the anomaly, DFT calculations were performed using
the {\sc Quantum ESPRESSO} (QE) \cite{GIANOZZI} code.  We used ultrasoft pseudopotentials (USPP)
treating the exchange-correlation energy functional  with a generalized gradient approximation (GGA) \cite{PERDEW}.
Since Bi has a strong spin-orbit coupling (SOC), we included SOC (through using fully relativistic USPP)
in the calculation of electronic properties. On the other hand, as the SOC does not change the phonon 
frequencies significantly,  vibrational properties were obtained without the inclusion of  SOC 
(using scalar relativistic pseudopotentials). We truncated the expansion of the Kohn-Sham wave functions
and the charge density in  plane wave basis with energy cut-offs of  60 Ry and 240 Ry,  respectively. 
To perform the Brillouin zone integrations, we used 12$\times$12$\times$2 uniform grid of
k-points. The discontinuity of the occupation numbers of the electronic states across the band gap
were smeared with the Fermi-Dirac distribution function. Phonon frequencies were obtained using  the linear response theory as implemented within the PH \cite{baroni2001phonons} package of QE.

In layered materials, stacking faults can occur with a relatively less energy penalty. We  explored the formation of
SFs as a possible mechanism of the anomaly as was done for Sb$_2$Te$_3$. Following Ref. [12], we introduced  stacking faults in the basal plane
of the hexagonal crystal structure (conventional unit cells) of Bi$_2$Se$_3$,
and constructed the pristine ($\vec{\mathbf{a}}_{_{0}}$, $\vec{\mathbf{b}}_{_{0}}$, $\vec{\mathbf{c}}_{_{0}}$)
and  faulted ($\vec{\mathbf{a}}_{_{\text{SF}}}=\vec{\mathbf{a}}_{_{0}}, \vec{\mathbf{b}}_{_{\text{SF}}}=\vec{\mathbf{b}}_{_{0}},
\vec{\mathbf{c}}_{_{\text{SF}}}= \vec{\mathbf{c}}_{_{0}} + \frac{1}{3}\vec{\mathbf{a}}_{_{0}} + \frac{2}{3}\vec{\mathbf{b}}_{_{0}}$)
configurations. In the above expression, $\vec{\mathbf{a}_{_{0}}}, \vec{\mathbf{b}_{_{0}}}$,
and $\vec{\mathbf{c}_{_{0}}}$  denote the lattice vectors for the hexagonal unit cell. We calculated the
free energy $\textrm{F}= \textrm{E}_{\textrm{tot}} + \textrm{F}_{\textrm{vib}}$ of these structures
as a function of temperature within a  harmonic approximation, where $\textrm{E}_{\textrm{tot}}$ 
is the total energy (calculated with SOC at 0 K)  and $\textrm{F}_{\textrm{vib}}$ is the vibrational contribution to the
free energy given by $\textrm{F}_{\textrm{vib}}=\frac{k_BT}{N_q}\sum_{iq}\textrm{log}[\textrm{2sinh}(\frac{\hbar\omega_{iq}}{2k_BT})]$,
$N_q$ being the total number of $q$ points in the Brillouin zone, $\omega_{iq}$ is the frequency 
of the $i$-th phonon mode with wave vector $q$ obtained using DFT linear response calculations.

\begin{figure}
\begin{center}
\includegraphics[width=1.0\linewidth]{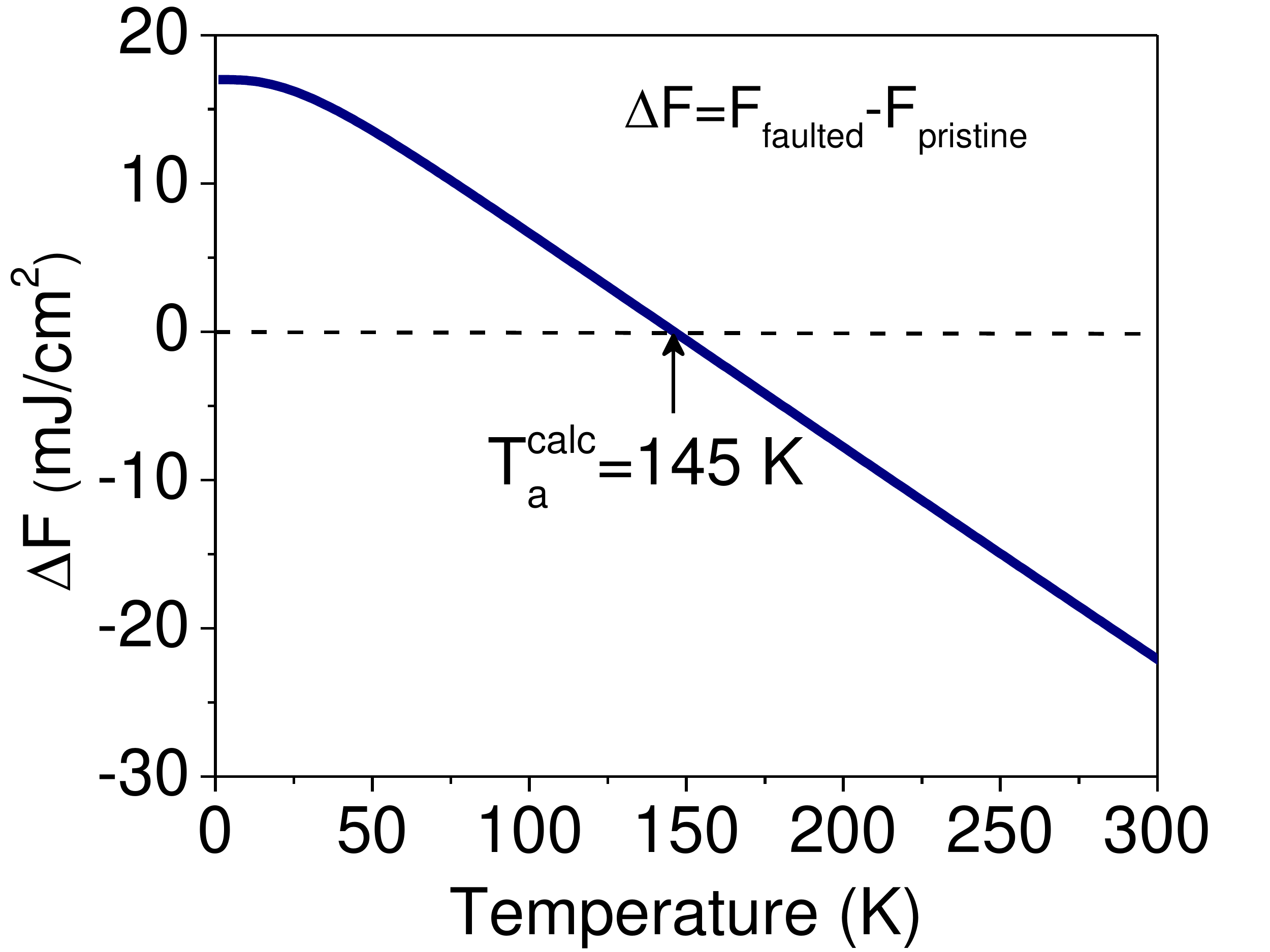}
\caption{Calculated free energy difference ($\Delta$F) of the faulted and pristine structures of Bi$_2$Se$_3$ as a function of temperature. }
\label{fig:FreeEnergyBi2Se3}
\end{center}
\end{figure}

After the optimization of the cell parameters (lattice vectors and the atomic positions) of the  pristine and faulted structures of Bi$_2$Se$_3$
using scalar relativistic USPPs, we relaxed the atomic positions of each configuration using fully
relativistic USPP to take into account the effect of SOC on the electronic states. The inclusion of
SOC significantly reduces  the total energy difference (at 0 K) between the pristine and faulted 
structures compared to the energy difference obtained without SOC. The energy difference without SOC was 72 meV/fu unit which reduced to 7 meV/fu after the inclusion of SOC.
 Hence, $\textrm{E}_{\textrm{tot}}$ 
(calculated with SOC) was added to $\textrm{F}_{\textrm{vib}}$ (obtained without SOC) to determine
 $\textrm{F}$ for each configuration of Bi$_2$Se$_3$. The calculated free energy difference
($\Delta \textrm{F}$ = F(faulted) -F(pristine))  decreases with temperature and  becomes negative at 145 K 
(see Fig. \ref{fig:FreeEnergyBi2Se3}) indicating a transition from the pristine to the faulted structure. This means that
 the structure of Bi$_2$Se$_3$ with infinitely extensive SFs will form and stabilize above T$_a^{calc}$=145 K.
 
We associate the theoretically estimate of T$_a^{calc}$=145 K  with the experimentally observed anomaly temperature $\text{T}_a=180$ K and attribute the origin of the observed anomalies in the optical constants ($n$ and $k$), sound velocity ($ v_{_{LA}}$) and Young's modulus of elasticity ($Y_{_{LA}}$) around 180 K to the formation of stacking faults near the anomaly temperature, consistent with the previous study \citep{Prakash}.  Our theoretical calculations on Bi$_2$Se$_3$ and earlier on Sb$_2$Te$_3$ establish firmly that the anomalous temperature dependence of the physical properties and associated vibrational anomalies are due to the ease of formation of stacking faults above a certain temperature. 

\section{\label{sec:level4}  Conclusion}
We have shown that the acoustic and optical phonons as well as the electronic relaxation dynamics in Bi$_2$Se$_3$ obtained using  ultrafast time resolved pump-probe spectroscopy, exhibits anomalous temperature dependence in the temperature range associated with an intriguing anomaly observed earlier 
in the thermal expansion. Using strain pulse propagation model, we extract the temperature dependence of the optical constants ($n$ and $k$), sound velocity ($v_{_{LA}}$) and the Young's modulus ($Y_{_{LA}}$),  which also show anomalous behavior around 180 K. These anomalies are  similar to that observed in Sb$_2$Te$_3$. The stacking fault mechanism  explaining the anomaly in Sb$_2$Te$_3$ is also the origin in Bi$_2$Se$_3$. Our DFT calculations show that inclusion of SOC is essential and plays an important role in reducing the total energy difference between the pristine and faulted structures of Bi$_2$Se$_3$.  The present work establishes that the anomalies in the thermal expansion, vibrational mode parameters, optical constants and their origin in the ease of forming stacking faults above a certain temperature are quite generic in nature in these layered 3D topological insulators. The presence of stacking faults at high temperatures should be considered in understanding their physical properties.

\begin{acknowledgments}
 AKS thanks Dr. Jayaraman for providing single crystals. AKS thanks DST for financial support.  GP thanks Council of Scientific and Industrial Research (CSIR) for SRF. AKS thanks DST for support through J.C. Bose National Fellowship. UVW acknowledges support from a J. C. Bose National Fellowship, a Sheikh Saqr Fellowship and IKST-Bangalore.
\end{acknowledgments}



%


\end{document}